# Alternative statistical methods for cytogenetic radiation biological dosimetry

Krzysztof Wojciech Fornalski [1,2]

[1] PGE EJ 1 Sp. z o.o., Technology and Operations Office, ul. Mysia 2, 00-496 Warszawa, Poland
[2] contract of specified task with Central Laboratory for Radiological Protection (CLOR), ul. Konwaliowa 7, 03-194 Warszawa, Poland
e-mail: krzysztof.fornalski@gmail.com, krzysztof.fornalski@gkpge.pl

**ABSTRACT**

The paper presents alternative statistical methods for biological dosimetry, such as the Bayesian and Monte Carlo method. The classical Gaussian and robust Bayesian fit algorithms for the linear, linear-quadratic as well as saturated and critical calibration curves are described. The Bayesian model selection algorithm for those curves is also presented. In addition, five methods of dose estimation for a mixed neutron and gamma irradiation field were described: two classical methods, two Bayesian methods and one Monte Carlo method. Bayesian methods were also enhanced and generalized for situations with many types of mixed radiation. All algorithms were presented in easy-to-use form, which can be applied to any computational programming language. The presented algorithm is universal, although it was originally dedicated to cytogenetic biological dosimetry of victims of a nuclear reactor accident.

## 1. INTRODUCTION

The aim of radiation biodosimetry by cytogenetics is to calculate doses and the associated confidence limits to exposed (or suspected exposed) persons after a radiation accident or incident. Calculating the dose absorbed in the human body is based on the observed chromosomal aberration (e.g. dicentrics) frequency in the lymphocytes of peripheral blood sampled from the exposed person (IAEA, 2001). This process requires the use of the fitted coefficients of the calibration dose-response curve that is produced by exposure of human blood *in vitro* to doses of the appropriate quality of radiation.

For accurate assessment of radiation doses and coefficients of the calibration dose response curve, a large number of mathematical and statistical methods need to be employed. A number of authors have suggested that the Bayesian statistics approach may be useful for analysis of cytogenetic data because it increases both the accuracy and quality assurance of radiation dose estimates (Brame & Groer, 2003; Ainsbury et al., 2013a, 2013b).

The objective of this paper is to present classical, Bayesian and Monte Carlo statistical methods which can be used in cytogenetic radiation biodosimetry for fitting the proper calibration curve, estimating the coefficients, selection of the dose-response model and estimating the dose and dose components from mixed radiation. The practical applications of those methods were presented in (Pacyniak et al., 2014).

## 2. CALIBRATION CURVES

In the production of an *in vitro* calibration curve for radiation dose assessment, the dose-response data obtained for different blood donors for the aberration induction in control and irradiated lymphocytes are collected and fitted to a linear or linear-quadratic model[1]. In this model, two DNA lesions in the two unduplicated chromosomes are required for producing chromosome interchanges, like dicentrics, and these lesions may arise from one or two independent ionization tracks (Kellerer & Rossi, 1974). Dicentrics produced by one track (single ionization) will have a frequency that is proportional to a linear function of dose ($aD$), whereas dicentrics induced by two tracks or photoelectric cascade will have a frequency proportional to the square of the dose ($bD^2$). In general, the shape of the dose-response relationship of chromosomal aberrations is strictly connected with the type of radiation and the way of ionization as well as with the linear energy transfer, LET.

### 2.1. Classical calibration curves

In the popular literature one can find the calibration curve of chromosomal aberrations for neutron irradiation, as (IAEA, 2001; Szłuińska et al., 2005):

$$Y_n(D_n) = Y_0 + \alpha\, D_n \qquad (1)$$

as well as for gamma radiation, as (IAEA, 2001; Szłuińska et al., 2005):

$$Y_g(D_g) = Y_0 + \beta\, D_g + \gamma\, D_g^{\,2} \qquad (2)$$

Equations (1)-(2) can be written in joined form (when the body is irradiated by mixed n+γ field) as:

$$Y_{n+g}(D_{n+g}) = Y_0 + a\, D_{n+g} + b\, D_{n+g}^2 \qquad (3)$$

Assuming that both radiation qualities are additive in the production of chromosomal damages, the dose-response relationship of chromosome aberrations may be described by:

$$Y_{n+g}(D_n, D_g) = Y_0 + \alpha\, D_n + \beta\, D_g + \gamma\, D_g^{\,2} \equiv y_f \qquad (4)$$

which is usually called a combined linear-quadratic equation for receiving the frequency of chromosomal aberrations $y_f$ after irradiation of mixed dose $D_n+D_g$. Parameters $Y_0$, $\alpha$, $\beta$ and $\gamma$ are usually found as results of the regression analysis. The parameter $y_f$ can be written as a ratio of $u/w$, where $u$ represents the number of chromosomal aberrations and $w$ – the number of cells.

### 2.2. Saturated calibration curves

All the equations (1)-(4) are widely used because of their simplicity and practicality. However, for high doses of several greys and more, linearity and parabolicality are not preserved (Sasaki, 2003). It is for this reason that cytogeneticists more experienced in the use of the mathematical methods can use a saturated version of equations (1)-(4). They are more general because equations do not trend to infinity ($\lim_{D\to\infty} Y^*(D) \neq \infty$), but the results of low and medium dose calculations are the same. Therefore, the linear function of eq. (1) can be replaced by a quasi-linear function:

$$Y_n^*(D_n) = (Y_{max} - Y_0) \cdot (1 - e^{-\alpha D_n}) + Y_0 \tag{5}$$

---

[1] In general mathematical terminology, the name „linear-quadratic" for equations (2)-(4) is incorrect; the correct name is "quadratic" or "parabolic"

Similarly, the linear-quadratic function of eq. (2) can be replaced by a sigmoid one:

$$Y_g^*(D_g) = (Y_{max} - Y_0) \cdot (1 - e^{-\beta D_g - \gamma D_g^2}) + Y_0 \tag{6}$$

The sigmoid function in eq. (6) can also be replaced by an Avrami sigmoid critical function, which is more adequate to radiation induced damages:

$$Y_g^{**}(D_g) = (Y_{max} - Y_0) \cdot (1 - e^{-aD_g^n}) + Y_0 \tag{7}$$

The $Y_{max}$ in equations (5)-(7) represents the maximum possible number of chromosomal aberrations per cell (in many cases one can assume simply $Y_{max}=1$), and $Y_0$ is the natural (non-radiation induced) level of spontaneous aberrations per cell. Therefore, $Y_0$ can correspond to the experiment average $Y_0 \approx 0.0005$ for dicentrics (Szłuińska et al., 2005).

### 2.3. Calibration curves for extreme doses

The irradiation by extreme doses, over a dozen greys, is rather an academic or laboratory case. However, in some situations saturated curves presented as eq. (5)-(7) are not accurate. When the dose increases over a certain critical point, the frequency of chromosomal aberrations can decrease due to cell death (Sasaki, 2003). In such a situation it is better to use the curve which is linear (linear-quadratic) in small and medium doses, saturates to critical point and decreases for highest doses, as in:

$$Y_n^\#(D_n) = (Y_{max} - Y_0) \cdot \alpha D_n \cdot e^{-\alpha D_n} + Y_0 \tag{8}$$

and

$$Y_g^\#(D_g) = (Y_{max} - Y_0) \cdot (\beta D_g + \gamma D_g^2) \cdot e^{-\beta D_g - \gamma D_g^2} + Y_0 \tag{9}$$

However, equations (5)-(9) can be used in special cases only, such as irradiation by high doses. For lower doses equations (1)-(4) are more common. Especially combined linear-quadratic eq. (4) will be used in the further part of this paper.

### 3. FITTING METHODS OF CALIBRATION CURVES

In the previous section the presentation covered the classical and modified calibration curves. In practice, the calibration curves are found due to the regression analysis methods of fitting the proper curve to specific experimental data. In the following section the classical Gaussian method is reminded and the robust Bayesian method is introduced.

### 3.1. Gaussian best fit

The method of curve to data fitting (regression analysis) used most often is a least squares method (Wolberg, 2005) connected with the general maximum likelihood method (IAEA, 2001). Just to remind, the maximum likelihood method provides the probability distribution (or probability density function, PDF) of e.g. a normal (Gaussian) distribution for all $N$ experimental data points ($D_i, E_i$) with vertical uncertainties $\sigma_{0i}$ each[2], where $Y_i$ represents a proposed model (calibration curve):

$$P = \prod_{i=1}^{N} P_i = \prod_{i=1}^{N} \frac{1}{\sigma_{0i}\sqrt{2\pi}} \exp\left[-\frac{(Y_i - E_i)^2}{2\sigma_{0i}^2}\right] \qquad (10)$$

The maximization of eq. (10) due to model's parameters will give the best fit of proper curve to the data points. However, for simplicity of analytical solutions the general maximum likelihood method allows to use the logarithm of $P$ and the sum instead of a product. Thus, the maximization of $P$ is equivalent to the minimization of the least square function:

$$\chi^2 = \sum_{i=1}^{N} \frac{(Y_i - E_i)^2}{\sigma_{0i}^2} \qquad (11)$$

which is called the least squares method. However, it was assumed that only the symmetrical vertical uncertainties ($\sigma_{0i}$) are taken into account (like presented ones in Fig. 1). In general situation with vertical ($\sigma_{y,i}$) and horizontal ($\sigma_{x,i}$) uncertainties, the denominator of eq. (11) can be replaced by $\sigma_{y,i}^2 + \left(\frac{dY_i}{dD_i}\right)^2 \sigma_{x,i}^2$. For simplicity, throughout the presented paper the horizontal uncertainties are assumed to be insignificant and $\sigma_{x,i}=0$.

The least squares method is simple and well accepted worldwide because the best fit to the experimental data points can easily be found. The method identifies the proper values of curve fitting parameters (equations (1)-(9)) very fast and effectively.

In the particular case of biological dosimetry the maximum likelihood method can be applied also to the Poisson distribution (Groer & Pereira, 1987; El-Sayyad, 1973). This method was used for neutron dosimetry for higher doses, where the background term $Y_0$ from eq. (1) can be omitted. The probability distribution of the slope $\alpha$ (eq. (1)) can be presented as:

$$P(\alpha) \propto \prod_{i=1}^{N} (\alpha D_i)^{u_i} \times \exp\left(-\sum_{i=1}^{N} w_i \alpha D_i\right) \qquad (12)$$

where $u_i$ represents the number of chromosomal aberrations in $i$-th sample of $w_i$ cells after receiving the dose $D_i$. Finding the maximum of eq. (12), one can easily calculate the best fit of $\alpha$ parameter.

All methods reminded above can be widely found in literature (Wolberg, 2005). However, those methods fail in the case of large scatter of experimental points and/or when at least one outlier point exists. In that case the more universal method is the robust Bayesian regression (fit) analysis described in the textbook by Sivia and Skilling (2006) and applied in author's papers (Fornalski & Dobrzyński, 2009, 2010a, 2010b, 2011; Fornalski et al., 2010).

---

[2] One has to note that in classical Gaussian regression all uncertainties of points, $\sigma_{0i}$, are the same

### 3.2. Bayesian best fit

The simple comparison between the robust Bayesian and least squares methods is presented in Fig. 1. One can clearly see that outliers make least squares method very misleading while Bayesian fit copes well and follows the main trend. This results from the fact that each *i*-th data point can be presented as a probability density function (PDF) composed of a proper Gaussian distribution (so called likelihood function, see eq. (10)) around its expected value as well as the prior function for its probability $\sigma_i$:

$$P_i = \int_{\sigma_{0i}}^{\infty} \frac{1}{\sigma_i \sqrt{2\pi}} \; e^{-\frac{(Y_i - E_i)^2}{2\sigma_i^2}} \times \frac{\sigma_{0i}}{\sigma_i^2} \, d\sigma_i \qquad (13)$$

The right-side prior function for $\sigma_i$ in eq. (13) assumes that the *i*-th analyzed probability $\sigma_i$ lies between the original one ($\sigma_{0i}$) and infinity. The procedure makes all outliers insignificant as input to the whole posterior probability distribution *P* for all *N* points, where, according to the maximum likelihood method described earlier (Fornalski et al., 2010) one can use a sum instead of a product:

$$P = \prod P_i \quad \Leftrightarrow \quad S = \sum \ln P_i \qquad (14)$$

where $P_i$ is a result of the integration of eq. (13) for single point *i*, see eq. (23).

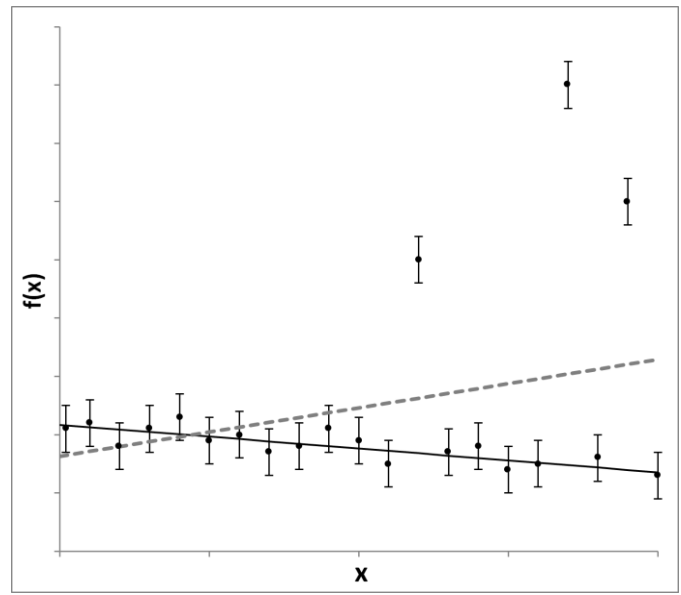

Figure 1. The example of robust Bayesian (black solid line) and least squares (grey dashed line) fits to some virtual experimental data with three outliers (outstanding points). More examples in (Fornalski et al., 2010).

After the differentiation of logarithmic probability *S* over all *n* fitting parameters $\lambda = \{\lambda_1, \lambda_2, ..., \lambda_n\}$, one can find the final and general form of a Bayesian fitting equation (Sivia & Skilling, 2006):

$$\frac{dS}{d\lambda} = \sum_{i=1}^{N} g_i(Y_i - E_i)\frac{dY_i}{d\lambda} \equiv 0 \tag{15}$$

where weights $g_i$ of the points are:

$$g_i = \frac{1}{(Y_i - E_i)^2}\left[2 - \frac{(Y_i - E_i)^2}{\sigma_{0i}^2} \times \frac{1}{\exp\left(\frac{(Y_i - E_i)^2}{2\sigma_{0i}^2}\right) - 1}\right] \tag{16}$$

The equation (15) can be implemented directly into the computational algorithm to find the best Bayesian fit to all $N$ experimental data points $(D_i, E_i)$ with vertical uncertainties $\sigma_{0i}$ each. The $Y_i$ means the theoretical shape of the best fit with $n$ fitting parameters, $Y_i(\lambda_1, \lambda_2, ..., \lambda_n)$, for example eq. (1)-(9). The result of the practical application (when eq. (1) for $Y_i$ is used) is presented in Fig. 1.

Exemplary solution of eq. (15), where $Y_i$ is given by eq. (4), is presented below.

In this special case of mixed n+γ radiation field (eq. (4)), the eq. (15) is described by the set of $n=4$ dependent equations ($\lambda = \{Y_0, \alpha, \beta, \gamma\}$):

$$\begin{cases} \sum_{i=1}^{N} g_i\left(Y_0 + \alpha D_{n,i} + \beta D_{g,i} + \gamma D_{g,i}^2 - E_i\right) = 0 \\ \sum_{i=1}^{N} g_i\left(Y_0 + \alpha D_{n,i} + \beta D_{g,i} + \gamma D_{g,i}^2 - E_i\right) D_{n,i} = 0 \\ \sum_{i=1}^{N} g_i\left(Y_0 + \alpha D_{n,i} + \beta D_{g,i} + \gamma D_{g,i}^2 - E_i\right) D_{g,i} = 0 \\ \sum_{i=1}^{N} g_i\left(Y_0 + \alpha D_{n,i} + \beta D_{g,i} + \gamma D_{g,i}^2 - E_i\right) D_{g,i}^2 = 0 \end{cases} \tag{17}$$

The next step is to solve the set of eq. (17) to find all four $\lambda$ parameters. The easiest way is to use the Cramer's rule and calculate determinants of *4x4* matrix:

$$W_0 = \det\begin{bmatrix} \sum_{i=1}^{N} g_i & \sum_{i=1}^{N} g_i D_{n,i} & \sum_{i=1}^{N} g_i D_{g,i} & \sum_{i=1}^{N} g_i D_{g,i}^2 \\ \sum_{i=1}^{N} g_i D_{n,i} & \sum_{i=1}^{N} g_i D_{n,i}^2 & \sum_{i=1}^{N} g_i D_{g,i} D_{n,i} & \sum_{i=1}^{N} g_i D_{g,i}^2 D_{n,i} \\ \sum_{i=1}^{N} g_i D_{g,i} & \sum_{i=1}^{N} g_i D_{n,i} D_{g,i} & \sum_{i=1}^{N} g_i D_{g,i}^2 & \sum_{i=1}^{N} g_i D_{g,i}^3 \\ \sum_{i=1}^{N} g_i D_{g,i}^2 & \sum_{i=1}^{N} g_i D_{n,i} D_{g,i}^2 & \sum_{i=1}^{N} g_i D_{g,i}^3 & \sum_{i=1}^{N} g_i D_{g,i}^4 \end{bmatrix} \tag{18a}$$

$$W_{Y_0} = \det\begin{bmatrix} \sum_{i=1}^{N} g_i E_i & \sum_{i=1}^{N} g_i D_{n,i} & \sum_{i=1}^{N} g_i D_{g,i} & \sum_{i=1}^{N} g_i D_{g,i}^2 \\ \sum_{i=1}^{N} g_i E_i D_{n,i} & \sum_{i=1}^{N} g_i D_{n,i}^2 & \sum_{i=1}^{N} g_i D_{g,i} D_{n,i} & \sum_{i=1}^{N} g_i D_{g,i}^2 D_{n,i} \\ \sum_{i=1}^{N} g_i E_i D_{g,i} & \sum_{i=1}^{N} g_i D_{n,i} D_{g,i} & \sum_{i=1}^{N} g_i D_{g,i}^2 & \sum_{i=1}^{N} g_i D_{g,i}^3 \\ \sum_{i=1}^{N} g_i E_i D_{g,i}^2 & \sum_{i=1}^{N} g_i D_{n,i} D_{g,i}^2 & \sum_{i=1}^{N} g_i D_{g,i}^3 & \sum_{i=1}^{N} g_i D_{g,i}^4 \end{bmatrix} \tag{18b}$$

$$W_\alpha = \det\begin{bmatrix} \sum_{i=1}^{N} g_i & \sum_{i=1}^{N} g_i E_i & \sum_{i=1}^{N} g_i D_{g,i} & \sum_{i=1}^{N} g_i D_{g,i}^2 \\ \sum_{i=1}^{N} g_i D_{n,i} & \sum_{i=1}^{N} g_i E_i D_{n,i} & \sum_{i=1}^{N} g_i D_{g,i} D_{n,i} & \sum_{i=1}^{N} g_i D_{g,i}^2 D_{n,i} \\ \sum_{i=1}^{N} g_i D_{g,i} & \sum_{i=1}^{N} g_i E_i D_{g,i} & \sum_{i=1}^{N} g_i D_{g,i}^2 & \sum_{i=1}^{N} g_i D_{g,i}^3 \\ \sum_{i=1}^{N} g_i D_{g,i}^2 & \sum_{i=1}^{N} g_i E_i D_{g,i}^2 & \sum_{i=1}^{N} g_i D_{g,i}^3 & \sum_{i=1}^{N} g_i D_{g,i}^4 \end{bmatrix} \tag{18c}$$

$$W_\beta = \det\begin{bmatrix} \sum_{i=1}^{N} g_i & \sum_{i=1}^{N} g_i D_{n,i} & \sum_{i=1}^{N} g_i E_i & \sum_{i=1}^{N} g_i D_{g,i}^2 \\ \sum_{i=1}^{N} g_i D_{n,i} & \sum_{i=1}^{N} g_i D_{n,i}^2 & \sum_{i=1}^{N} g_i E_i D_{n,i} & \sum_{i=1}^{N} g_i D_{g,i}^2 D_{n,i} \\ \sum_{i=1}^{N} g_i D_{g,i} & \sum_{i=1}^{N} g_i D_{n,i} D_{g,i} & \sum_{i=1}^{N} g_i E_i D_{g,i} & \sum_{i=1}^{N} g_i D_{g,i}^3 \\ \sum_{i=1}^{N} g_i D_{g,i}^2 & \sum_{i=1}^{N} g_i D_{n,i} D_{g,i}^2 & \sum_{i=1}^{N} g_i E_i D_{g,i}^2 & \sum_{i=1}^{N} g_i D_{g,i}^4 \end{bmatrix} \tag{18d}$$

$$W_\gamma = \det\begin{bmatrix} \sum_{i=1}^{N} g_i & \sum_{i=1}^{N} g_i D_{n,i} & \sum_{i=1}^{N} g_i D_{g,i} & \sum_{i=1}^{N} g_i E_i \\ \sum_{i=1}^{N} g_i D_{n,i} & \sum_{i=1}^{N} g_i D_{n,i}^2 & \sum_{i=1}^{N} g_i D_{g,i} D_{n,i} & \sum_{i=1}^{N} g_i E_i D_{n,i} \\ \sum_{i=1}^{N} g_i D_{g,i} & \sum_{i=1}^{N} g_i D_{n,i} D_{g,i} & \sum_{i=1}^{N} g_i D_{g,i}^2 & \sum_{i=1}^{N} g_i E_i D_{g,i} \\ \sum_{i=1}^{N} g_i D_{g,i}^2 & \sum_{i=1}^{N} g_i D_{n,i} D_{g,i}^2 & \sum_{i=1}^{N} g_i D_{g,i}^3 & \sum_{i=1}^{N} g_i E_i D_{g,i}^2 \end{bmatrix} \tag{18e}$$

The request fitting parameters $\lambda$ can be calculated[3] as $Y_0 = W_{Y_0}/W_0$, $\alpha = W_\alpha/W_0$, $\beta = W_\beta/W_0$ and $\gamma = W_\gamma/W_0$. However, such parameters have to be iteratively calculated using a computational algorithm[4], because all $\lambda$ are also implicit in $g_i(\lambda)$ (eq. (16)). Finally, the results are all $n=4$ fitting parameters $\lambda=\{Y_0, \alpha, \beta, \gamma\}$ of the curve $Y_{n+g}(D)$ given by eq. (4).

### 3.3. Uncertainties

The fitting parameters $\lambda=\{\lambda_1, \lambda_2, ..., \lambda_n\}$ found thanks to the eq. (15) have their own uncertainties $\sigma_\lambda=\{\sigma_1, \sigma_2, ..., \sigma_n\}$, which can be estimated using the Hessian matrix, $H$ (Fornalski & Dobrzyński, 2009, 2010b). Thus, the uncertainty of exemplary $\lambda_n$ parameter can be calculated from $n$-th variance, which is equal to an $h_{n,n}$ element of main diagonal from invertible Hessian matrix, $H^{-1}$:

$$\sigma_n = \sqrt{\frac{(-1)^{2n} H_{n,n}}{\det H}} \tag{19}$$

However, the calculation of eq. (19) is rather difficult (but recommended) for high values of $n$, so $\sigma_n$ can be sometimes approximated using the Cramér-Rao theorem[5] (see also eq. (42)):

$$\sigma_n \geq \frac{1}{\sqrt{\omega_n}} \tag{20}$$

where

$$\omega_n = \frac{d^2 S}{d\lambda_n^2} = -\sum_{i=1}^{N} \left[\xi_i (Y_i - E_i)^2 - g_i\right] \left(\frac{dY_i}{d\lambda_n}\right)^2 \tag{21}$$

and

$$\xi_i = \frac{1}{P_i} \frac{4}{R_i^6} \left[2 - \left(2 + \frac{R_i^2}{\sigma_{0i}^2} + \frac{R_i^4}{4\sigma_{0i}^4}\right) \exp\left(\frac{-R_i^2}{2\sigma_{0i}^2}\right)\right] - g_i^2 \tag{22}$$

The residuals $R_i = Y_i - E_i$ can also be used for all previous equations just for simplicity. Additionally, the $P_i$ function is the result of an integral from eq. (13):

$$P_i = \frac{\sigma_{0i}}{R_i^2 \sqrt{2\pi}} \left[1 - \exp\left(\frac{-R_i^2}{2\sigma_{0i}^2}\right)\right] \tag{23}$$

Using eq. (20)-(23) one can calculate the lower bound of uncertainty $\sigma_n$ of fitting parameter $\lambda_n$. However, the eq. (19) should be used to have exact values of uncertainties.

---

[3] sometimes, just for simplicity, the variable $D_{n,i}$ can be presented as a function of $D_{g,i}$, using eq. (30)
[4] The actual value of $\lambda$ is putting into the $g_i(\lambda)$, the next iteration will give more precise $\lambda'$, which is putting into the $g_i(\lambda')$ etc.
[5] Cramér-Rao theorem gives lower bounds of the variances of the estimators, that are just the elements of the diagonal of the inverse of the Fisher information matrix. These lower bounds are asymptotically attained by the variances of maximum likelihood estimators, and the inverse of the Fisher information matrix can be estimated from the inverse of the Hessian matrix, $H^{-1}$, of the log-likelihood function evaluated at the maximum likelihood estimates

### *3.4. Model selection*

The Bayesian analysis allows also the possibility of relative estimation of the proposed model reliability[6]. Thanks to that one can check which fitted function, for example from eq. (1)-(9), is the best solution for existed data points (Fornalski & Dobrzyński, 2010b, 2011).

The Bayes theorem connects the probabilities of *P(Model|Data) ~ P(Data|Model)*, which can be used to estimate the relative reliability of two models, *M*, in the case of the same data, *D*. To do so, it is necessary to calculate the reliability function for the model *M* with the fitting parameter *λ*, using the marginalization procedure:

$$P(M|D) \propto P(D|M) = \int P(D,\lambda|M)\, d\lambda = \int P(D|\lambda, M) \times P(\lambda|M)\, d\lambda \qquad (24)$$

The *P(D|λ,M)* corresponds to the likelihood function, represented by the Gaussian distribution around the expected value $\lambda_0 \pm \sigma_\lambda$ with maximum probability of likelihood function equals *P(D|λ$_0$,M)*. The prior probability *P(λ|M)* can be assumed as a uniform distribution $U(\lambda_{min}, \lambda_{max})$. Because such form of *P(λ|M)* is independent of *λ*, the integral (24) can be written as:

$$P(D|M) = \frac{1}{\lambda_{max} - \lambda_{min}} \int_{\lambda_{min}}^{\lambda_{max}} P(D|\lambda_0, M)\, e^{-\frac{(\lambda - \lambda_0)^2}{2\sigma_\lambda^2}}\, d\lambda \qquad (25)$$

The result of the integral (25) can be approximated by $P(D|\lambda_0, M)\, \sigma_\lambda \sqrt{2\pi}$ because "*the sharp cut-offs at $\lambda_{min}$ and $\lambda_{max}$ do not cause a significant truncation of the Gaussian*" probability distribution from eq. (25) (Sivia & Skilling, 2006). Because $\lambda_0$ corresponds to the parameter found by the robust Bayesian best fit method for model *M*, the maximum value of likelihood function *P(D|λ$_0$,M)* can be replaced by the set of $P_i$ given by eq. (13) or (23) and the final form of the reliability function can be approximated by (Fornalski & Dobrzyński, 2011):

$$P(M|D) \propto P(D|M) \approx \sum P_i \times \frac{\sigma_\lambda \sqrt{2\pi}}{\lambda_{max} - \lambda_{min}} \qquad (26)$$

The right-hand term in eq. (26) is called an Ockham factor, which prevents the use of overcomplicated models. Equation (26) corresponds to the situation, where model *M* have only one (*n=1*) fitting parameter, $\lambda_0 \pm \sigma_\lambda$. In the case of zero parameters (*n=0*) the Ockham factor equals 1 and model *M* is just a constant value ($Y_i$=const). In the case of *n* fitting parameters $\lambda = \{\lambda_1, \lambda_2, ..., \lambda_n\}$ with their estimated uncertainties $\sigma_\lambda = \{\sigma_1, \sigma_2, ..., \sigma_n\}$, the most general form of eq. (26) can be presented as (Fornalski & Dobrzyński, 2011):

$$P(M|D) \propto \sum_{i=1}^{N} \frac{1}{(Y_{M,i} - E_i)^2}\left[1 - \exp\left(-\frac{(Y_{M,i} - E_i)^2}{2\sigma_{0i}^2}\right)\right] \times \prod_{\lambda=1}^{n} \frac{\sigma_\lambda \sqrt{2\pi}}{\lambda_{max} - \lambda_{min}} \qquad (27)$$

It is important to remember that *N* represents the number of experimental points ($D_i, E_i$) with vertical uncertainties $\sigma_{0i}$ each, to which model *M* (the *Y(D)* relationship) is fitted using *n* fitting parameters $\lambda \pm \sigma_\lambda$. The most problematic are the values $\lambda_{min}$ and $\lambda_{max}$, which can be calculated independently or taken arbitrary for all *λ*. In the case of the arbitrary method, they can be taken as minimum/maximum possible values of the considered parameter *λ* using the largest span that can be tolerated by the data. In order not to extend the range of *λ* in the case of a huge scatter of data, not more than three points are allowed to lie outside the range (Fornalski & Dobrzyński, 2011).

---

[6] Thus, the classical methods of model selection (like AIC) or other Bayesian ones (like BIC) will be omitted here

One can also find a non-arbitrary way to calculate both parameters $\lambda_{min}$ and $\lambda_{max}$. Such a method should be independent to Bayesian regression, so one can propose for example $\lambda_{min} = \lambda - k\sigma_\chi$ and $\lambda_{max} = \lambda + k\sigma_\chi$, where $\sigma_\chi$ means the standard deviation of Gaussian regression (see eq. (10)) and $k$ the degree of belief (e.g. $k \equiv 2$).

The eq. (27) can be applied directly in the computational algorithm (as well as the Excel's formula) to calculate reliability functions $P(M=Y(D)|D)$ for each model, like eq. (1)-(9). After that one can calculate the relative value of each two models, say $A$ and $B$, to check which of them is more likely to describe the data:

$$W_M = \frac{P(M=A|D)}{P(M=B|D)} \tag{28}$$

When $W_M$ is greater than 1, model $A$ wins over $B$. When $W_M \approx 1$, both models have the same degree of belief. In general, $W_M$ can quantify the preference of one model with respect to the other one. In practice, the real values of $W_M$ may show that the plausibility of models can differ by orders of magnitude (Fornalski & Dobrzyński, 2011).

### 4. DOSE ESTIMATION METHODS

Dose estimation methods in biological dosimetry using e.g. Bayesian statistics can be found in the literature (Brame & Groer, 2003; Ainsbury et al., 2013a, 2013b). However, the methods presented below are more wide and can be treated as a significant extension of the general biodosimetry statistics.

Having the fitting parameters of the calibration curves estimated by Gaussian or Bayesian method (previous section), one can use $Y(D)$ functions to estimate the dose ($D$) and/or the frequency of chromosome aberrations ($Y$) after accidental exposure of people to gamma, neutron or to mixed n+γ radiation (for example after nuclear reactor accident). Because such mixed radiation is composed of particles having different biological effects, there is a strong need to calculate not only the total absorbed dose but also components of the total dose (here $D_n$ and $D_g$). From the observed chromosome aberration frequency in the lymphocytes irradiated with mixed n+γ radiation it is not possible to discriminate between those aberrations due to γ-rays and those due to neutrons. However, in the case where a physical estimate of the neutron to gamma absorbed doses ratio is known:

$$\rho = \frac{D_n}{D_g} \tag{29}$$

it is possible to estimate the separate neutron and γ-ray doses by iterative method (IAEA, 2001; Szłuińska et al., 2005), which will not be described here. However, if a physical estimate of neutron and photon components of the absorbed dose is not available, some alternative statistical methods can be applied.

In further considerations, the eq. (4) for chromosomal aberration frequency, $Y$, will be taken into account.

### *4.1. Prior functions*

Assuming that calibration curve $Y_{n+g}(D_n,D_g)$ is measured precisely, the most important problem is connected with the precise measurement of neutron to gamma absorbed doses ratio (eg. (29)). The easiest case is when the ratio is known exactly or with an uncertainty $\rho \pm \sigma_\rho$. However, using $\rho$ which can vary from zero to infinity is not practical. The $\theta$ parameter normalized to the range of [0,1] is recommended as (Brame & Groer, 2003)

$$\theta = \frac{D_g}{D_g + D_n} = \frac{1}{\rho + 1} \tag{30}$$

Parameter $\theta$ defined by eq. (30) corresponds to the contribution of gamma dose in the total dose and is more practical in use than $\rho$. However, the problems for $\rho$ appear also for $\theta$ in the context of precise dose measurement.

When $\theta$ (or $\rho$) is not precisely known, one can use the PDF to estimate the most probable value of $\theta$ (or $\rho$). Such a PDF can be called an prior probability p($\theta$) or p($\rho$) and for the most common cases it can be approximated by Gaussian distribution[7] for $\theta$ (or $\rho$) with standard deviation of $\sigma_\theta$ (or $\sigma_\rho$), for example:

$$p(\theta) = \frac{1}{\sqrt{2\pi}\sigma_\theta} \exp\left[-\frac{(\theta - \hat{\theta})^2}{2\sigma_\theta^2}\right] \tag{31}$$

where $\hat{\theta}$ represents the expected value with uncertainty $\sigma_\theta$.

For normalized values [0,1] one can use analogical Gaussian prior for $\hat{\rho} \pm \sigma_\rho$ but transformed into $\theta$ coordinates (Brame & Groer, 2003):

$$p(\theta) = \frac{1}{\sqrt{2\pi}\sigma_\rho \theta^2} \exp\left[\frac{-1}{2\sigma_\rho^2}\left(\left(\frac{1}{\theta} - 1\right) - \hat{\rho}\right)^2\right] \tag{32}$$

The main assumption for both prior functions of eq. (31) and (32) is that information about the expected value and its uncertainty is necessary (so called informative priors[8]). However, in some cases such data does not exist and one the only information available is that there is a mixed n+γ radiation field with unknown $\theta$. In that case one can assume that there is approximately an even proportion of neutron and gamma doses, which correspond to the uninformative prior (from Beta distribution):

$$p(\theta) = \theta - \theta^2 \tag{33}$$

or

$$p(\theta) = const \tag{34}$$

when one can say nothing about the gamma to neutron relation. The prior (34) can also be approximated by the uniform probability distribution within an assumed range, $p(\theta)=U(\theta_{min},\theta_{max})$. However, the use of prior (34) is not always possible and rather gives the least precise results whenever used. It is the reason why one should always try to assess the potential p($\theta$) shape, which is usually possible.

---

[7] In some cases the proper informative prior distribution can use Beta or Log-normal distribution
[8] When our knowledge about $\theta$ is more and more accurate, informative priors would asymptotically aim towards the Kronecker delta (see http://en.wikipedia.org/wiki/Kronecker_delta)

Now, having a sufficient information about $\theta$, one can try to estimate the n+γ doses and chromosomal aberration frequencies, especially for dicentrics. Five alternative statistical methods are proposed below.

### *4.2. Classical method*

In the classical method the contribution of gamma dose in the total dose, $\theta$, is precisely known and $\theta$ is given as an exact value. Thus no prior function for $\theta$ is needed.

The most common approach is an iterative algorithm, precisely described in (IAEA, 2001; Szłuińska et al., 2005). In this approach the doses and chromosomal aberration frequencies are estimated in iterative way from eq. (4) and (29). However, this algorithm is usually laborious and requires many repetitions (iterations) to obtain final results.

To automate this method, one can use a system of two equations: eq. (30) and exemplary eq. (4):

$$\begin{cases} y_f = Y_0 + \alpha D_n + \beta D_g + \gamma D_g^2 \\ \theta = \frac{D_g}{D_g + D_n} \end{cases} \quad (35)$$

to calculate both doses, $D_g$ and $D_n$, in the function of $\theta$:

$$\begin{cases} D_g(\theta) = \frac{\sqrt{\left(\alpha\frac{1-\theta}{\theta}+\beta\right)^2 + 4\gamma(y_f - Y_0)} - \left(\alpha\frac{1-\theta}{\theta}+\beta\right)}{2\gamma} \\ D_n(\theta) = \frac{1-\theta}{\theta} D_g(\theta) \end{cases} \quad (36)$$

in the special case of $Y(D)$ given by eq. (4). It is assumed, that all constants ($\alpha$, $\beta$, $\gamma$, $y_f$, $Y_0$ and $\theta$) are precisely known due to the experimental data with proper uncertainties (see previous section). Treating $\theta$ as a variable, one can present a set of eq. (36) in Fig. 2.

Finally, the uncertainties of dose estimations, $\sigma_D$, in the presented method can be calculated using the sum of independent finite increments method:

$$\sigma_{D_x} \equiv \sum_{j=1}^{j_{max}} \left|\frac{\partial D_x}{\partial \varepsilon_j}\right| \delta\varepsilon_j \quad (37)$$

where $x=\{g,n\}$ and $j_{max}$ represents the total number of parameters $\varepsilon_j \pm \delta\varepsilon_j$. Assuming the $Y(D)$ given by eq. (4), parameters $\varepsilon_j=\{\alpha,\beta,\gamma,y_f,Y_0,\theta\}$ and $\delta\varepsilon_j=\{\sigma_\alpha,\sigma_\beta,\sigma_\gamma,\sigma_{yf},\sigma_{Y0},\sigma_\theta\}$ can be found experimentally (see previous section). The eq. (37) will give slightly higher values than in the exact differential method.

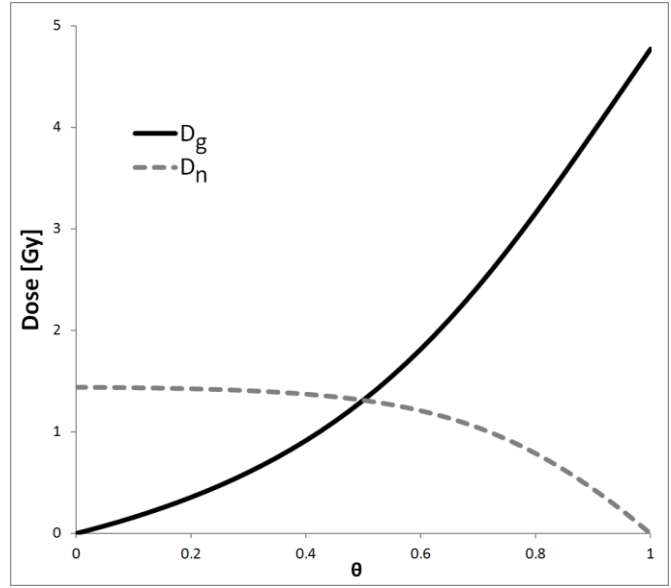

Figure 2. The graphical representation of eq. (36) for exemplary experimental data α=0.832, β=0.0164, γ=0.0492, $y_f$=1.2 and $Y_0$=0.0005 (IAEA, 2001).

### *4.3. Enhanced classical method (quasi-Bayesian method)*

To enhance the classical method for the possibility, that $\theta$ is given not by the value but by the prior function (PDF), one needs to transform classical relationships into the probability distributions[9]. In practice, the set of eq. (35) needs to be solved to find distributions of $\theta$ in the example case of eq. (4):

$$\begin{cases} \theta_g(D_g) = \dfrac{D_g}{D_g + \frac{1}{\alpha}(y_f - Y_0 - \beta D_g - \gamma D_g^{\,2})} \\ \theta_n(D_n) = \dfrac{\sqrt{\beta^2 - 4\gamma(Y_0 + \alpha D_n - y_f)} - \beta}{\sqrt{\beta^2 - 4\gamma(Y_0 + \alpha D_n - y_f)} - \beta + 2\gamma D_n} \end{cases} \quad (38)$$

The two different designations of $\theta$ result from the fact, that the parameter is not precisely described by a value but by a prior probability function. Thus, the probability distribution of the dose can be written as:

$$p(\theta)\,\theta'(D) = p(\theta_x(D_x))\,\theta_x'(D_x) \equiv const \cdot P(D_x) \quad (39)$$

where $x=\{g,n\}$. For example eq. (39) can be presented with exemplary prior from eq. (33) as a system of two probability distributions:

$$\begin{cases} P(D_g) \propto \theta_g(D_g) - \theta_g(D_g)^2 \\ P(D_n) \propto \theta_n(D_n) - \theta_n(D_n)^2 \end{cases} \quad (40)$$

presented in Fig. 3.

---

[9] That way the method can also be called the *Bayesian-frequentist hybrid* method.

Finally, the estimation of the dose $D_x$ can be found from the maximum of the distribution (39), taken directly from the graph (like exemplary Fig. 3 and eq. (40)) or calculated from the first derivate equation:

$$\frac{d\,P(D_x)}{dD_x} = 0 \qquad (41)$$

The exemplary results shown in Fig. 3 correspond well with the *Classical method* calculations of eq. (36) for precisely known $\theta=0.5$ (equals to the expected value of prior (33)).

The uncertainties of dose estimations, $\sigma_{Dx}$, in the presented method can be assessed from the shape of distributions or calculated using the Cramér-Rao theorem:

$$\sigma_{D_x} \geq \frac{1}{\sqrt{\left|\frac{\partial^2 \ln P}{\partial D_x^2}\right|}} \qquad (42)$$

where *ln(P)* is a natural logarithm of *P(D$_x$)* due to the maximal likelihood method.

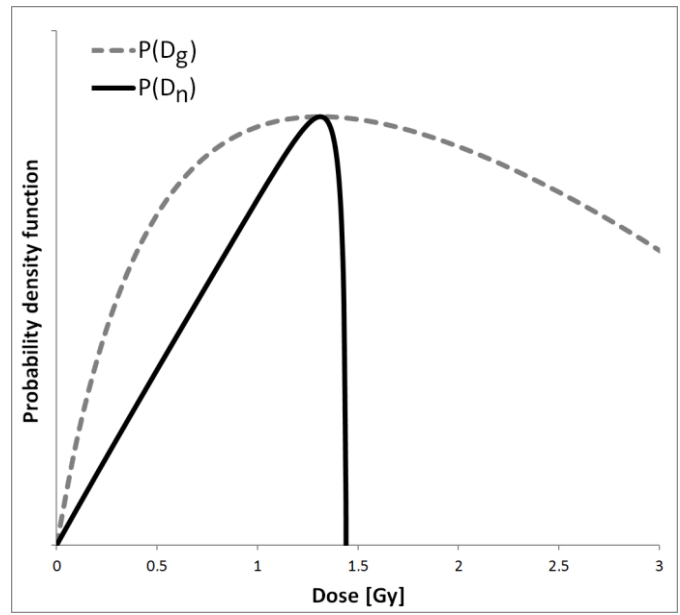

Figure 3. The graphical representation of eq. (40) for exemplary experimental data α=0.832, β=0.0164, γ=0.0492, y$_f$=1.2 and Y$_0$=0.0005 (IAEA, 2001) with prior given by eq. (33).

### 4.4. Simplified Bayesian method

The Bayesian reasoning can be reduced to the simple probabilities equation:

$$POSTERIOR\ PROB. = LIKELIHOOD\ PROB. \times PRIOR\ PROB. \qquad (43)$$

This type of equation was formerly used in the case of eq. (13) and (25). In the case of dose estimation, the selection of prior probability function was discussed for equations (31)-(34). The likelihood probability function has to be found using biophysical basis.

In general, there are $w$ cells and $u$ events (here: chromosomal aberrations). The expected number of events equals $w \cdot y_f$, so one can use the Poisson statistics for likelihood function as:

$$L = \frac{(w\, y_f)^u\, e^{-w\, y_f}}{u!} \qquad (44)$$

Taking equations (4) and (30) for eq. (44), one can calculate the likelihood functions for both n+γ doses:

$$\begin{cases} L(D_g|\theta) = \dfrac{\left[w\,(Y_0+\alpha\frac{1-\theta}{\theta}D_g+\beta D_g+\gamma D_g^2)\right]^u}{u!} \times e^{-w\,(Y_0+\alpha\frac{1-\theta}{\theta}D_g+\beta D_g+\gamma D_g^2)} \\ L(D_n|\theta) = \dfrac{\left[w\,(Y_0+\alpha D_n+\beta\frac{\theta}{1-\theta}D_n+\gamma(\frac{\theta}{1-\theta}D_n)^2)\right]^u}{u!} \times e^{-w\,(Y_0+\alpha D_n+\beta\frac{\theta}{1-\theta}D_n+\gamma(\frac{\theta}{1-\theta}D_n)^2)} \end{cases} \qquad (45)$$

Having the likelihood functions (45) and the assumed prior function $p(\theta)$, one can find the posterior probability – the probability distribution of dose :

$$P(D_x) = \int_0^1 L(D_x|\theta)\, p(\theta) d\theta \qquad (46)$$

where $x=\{g,n\}$. Similarly to the case of *Enhanced classical method*, one can find the estimation of dose $D_x$ from the maximum of the curve (46) or calculating the first derivate equation, given by the same eq. (41). The result of eq. (41) in the context of eq. (46) gives the searched estimated values of doses. However, for some priors, due to the maximum likelihood method, the natural logarithm of $P$ is much easier in analytical calculations.

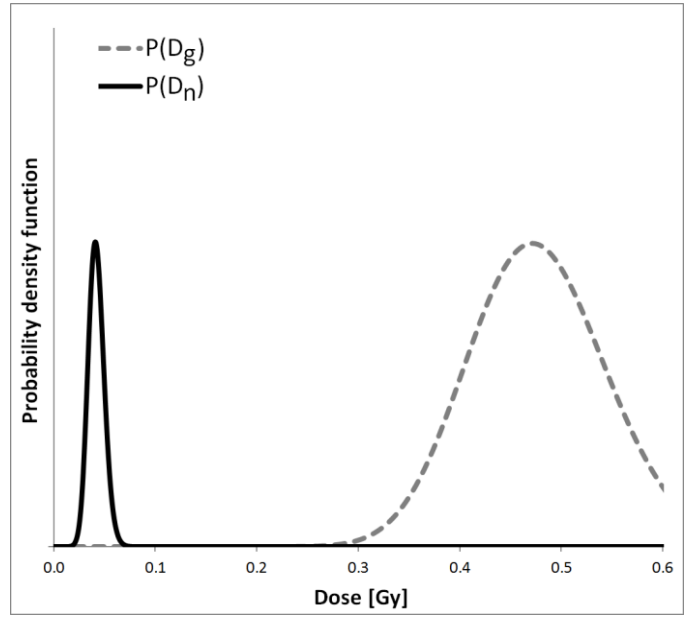

Figure 4. The application of Bayesian method (eq. (46)) to real experimental data, where $w=1000$, $u=33$, $\alpha=0.354$, $\beta=0.0119$, $\gamma=0.0557$ and $Y_0=0.0005$ (Pacyniak et al., 2014). The prior (32) with $\hat{\theta} = 0.92$ was used.

K.W. Fornalski: Alternative statistical methods for cytogenetic radiation biological dosimetry. arXiv.org, 2014

However, the analytical solution of integral (46) is often difficult (or impossible) to obtain. It is then possible to use computational methods of integration, such as the Monte Carlo integration method or iterative integration method. The latter method is quick, however it can cause some instabilities and fluctuations of the final shape of the *P(D_x)* curve due to large values of *u* variable[10].

Finally, the uncertainty of estimated value of dose, $\sigma_{Dx}$, can be approximated using the eq. (42). Fig. 4 contains the application of presented method to the real experimental data.

Additionally, the method presented above can also be simplified with the use of the exact value of $\theta$ instead of prior function, p($\theta$). In that way one should put the known $\theta$ directly into eq. (45) and analyze the distribution of *L(D_x)* to find its maximum for *D_x* estimation.

### *4.5. Full Bayesian method*

The complete Bayesian approach of dose estimation was proposed by Dr. R.S. Brame and Prof. P.G. Groer (Brame & Groer, 2003). They assumed that not only $\theta$ parameter is given by a prior function (they used p($\theta$) as eq. (32)), but also $\alpha$, $\beta$ and $\gamma$ are given by a certain probability distribution. In general, prior functions for $\alpha$, $\beta$ and $\gamma$ are given by a proper posterior probability distributions (eq. (14)) for fitted parameters of calibration curve. Another words: priors p($\lambda$) are the results of robust Bayesian fitting method (see previous section). However, Brame and Groer (2003) assumed for simplicity that those priors can be approximated by the Gamma distribution (the logarithm of Gamma distribution has a simple form for analysis, e.g. for maximum likelihood method):

$$p(\lambda) = \lambda^{k-1} \frac{z^k}{\Gamma(k)} \exp(-z\lambda) \qquad (47)$$

where $\lambda=\{\alpha,\beta,\gamma\}$ and $\Gamma$ is a gamma function. Parameters *k* and *z* are the shape and scale parameters, respectively. Having such an assumption, one can write the PDF of dose as

$$P(D_x) \propto \int \int \int \int L(D_x|\alpha,\beta,\gamma,\theta)\, p(\alpha)\, p(\beta)\, p(\gamma)\, p(\theta)\, d\alpha\, d\beta\, d\gamma\, d\theta \qquad (48)$$

The likelihood function, *L*, is also given by the Poisson distribution (44). When parameters $\lambda$ are given in the classical way with some standard deviation, priors p($\lambda$) can be introduced by a Gaussian distribution, like eq. (31). One has to note that Brame and Groer (2003) did not use the prior function for *Y_0* parameter (see eq. (4)), but generally this should also be included.

Finally, the eq. (48) can be calculated using computational or analytical methods (e.g. maximum likelihood method), similarly to eq. (46).

Details and results of eq. (48) are presented in the paper by Brame and Groer (2003).

---

[10] Large *u!* gives zero or infinity in some programs or codes, which can cause wrong results

### 4.6. Monte Carlo method

The Monte Carlo method is a computational combination of classical iterative method and *Enhanced classical method (quasi-Bayesian method)*. This method requires a dedicated computer program[11] and an implementation of the algorithm presented in Fig. 5.

In general, the algorithm is composed of two major loops: one over *w* cells and the second over *K* iterations. The loop over *K* is dedicated to obtain average results of *K* independent simulations. Stable results are obtained when *K* equals several hundred of Monte Carlo repetitions (thus, proper uncertainties can be estimated as a standard deviation of the results' normal distribution).

In the single *i*-th simulation, the current cell is randomized from all *w* cells. In such a case, the actual value of $\theta$ is taken. It can be an exact value of $\theta_0$ when such a value is available or is randomized from the probability distribution of p($\theta$). In the latter case, the actual value of $\theta_j$ is taken from the classical Monte Carlo randomization, when the maximal value of p($\theta$) equals $p_{max}$ (for normalized probability $p_{max}=1$).

The actual value of $\theta$ is used to select the type of interaction: neutron or gamma. For the damage from neutron interaction, the variable $u_n$ increases by 1. For the damage from gamma – $u_g$ increases by 1. The DNA damage from radiation is a primary cause which can be used to calculate the absorbed dose from $u_x$ damages ($x=\{g,n\}$):

$$D_x = f(u_x) \approx const \cdot u_x \qquad (49)$$

The function *f* from eq. (49) can be linear, quadratic or generally polynomial or saturated. For simple use of the algorithm, one can assume the linear relationship with the slope *const ≈ 0.012* calculated from comparison between Monte Carlo results and real data from (IAEA, 2001). Thus, having the values of $D_n$ and $D_g$, one can calculate proper chromosomal aberration frequencies $Y_n$ and $Y_g$, e.g. from eq. (1) and (2). The loop over *w* cells continues when the actual sum $Y_n+Y_g$ is lower than assumed $y_f$. The whole algorithm is presented graphically in Fig. 5.

Owing to the presented Monte Carlo method one can tests all statistical aspects of such a virtual irradiation of cells. For example one can add the information about each damage from irradiation, $u_x$, in the dedicated table *T* of all *w* cells. The exact number of damages can then be correlated with dose and chromosomal aberrations in single cells. Also one can obtain many useful distributions of cell parameters, such as Poisson distribution of damage in cells. It provides further statistical studies on virtual irradiation of a group of cells.

## 5. GENERALIZATION

Methods presented in previous sections were introduced for two types of radiation, especially gamma and neutron. However, generally the problem can be enhanced to the situation, where the potential victim is irradiated by many different types of radiation, as during a severe nuclear accident or space travel.

---

[11] The mentioned computational program was created by the author using C++ language

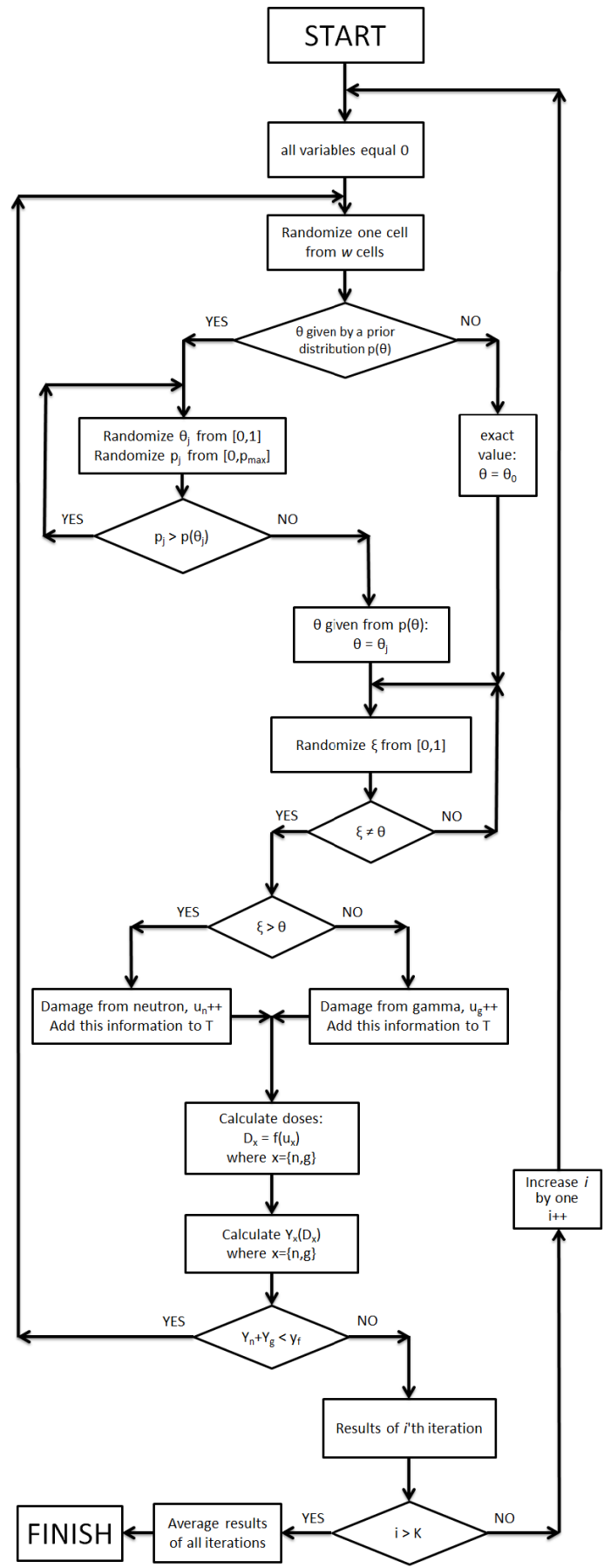

Figure 5. The detailed scheme of the Monte Carlo method's algorithm.

The generalized calibration curve, combined Bayesian and Gaussian regression fit as well as generalized Bayesian dose estimation methods are introduced below.

### 5.1. Generalized calibration curve

The calibration curves from eq. (1) and (2) can be written in a more general form, as a polynomial

$$Y(D) = Y_0 + \lambda_1 D + \lambda_2 D^2 + \ldots + \lambda_n D^n \qquad (50)$$

In the case of many types of radiation, like eq. (4), the eq. (50) can be presented as

$$Y(D_{total}) = Y_0 + \sum_{i=1}^{R} \sum_{j=1}^{n} \lambda_{i,j} D_i^j \qquad (51)$$

where *n* is a degree of a *i*-th polynomial and *R* is a number of radiation types.

A more complicated form of generalized *Y(D)* is expected for saturated and critical equations (5)-(9). However, one can find a single general equation, which is common for all presented before, namely eq. (1)-(9) and (50)-(51):

$$Y(D_{total}) = Y_0 + (Y_{max} - Y_0) \sum_{i=1}^{R} \left[ \sum_{j=0}^{n} \left( \lambda_{i,j}^{(a)} + \lambda_{i,j}^{(b)} D_i^j e^{-\sum_{k=1}^{K} \lambda_{i,j,k}^{(c)} D_i^{m_{i,j,k}}} \right) \right] \qquad (52)$$

For *λ(a)=λ(b)=0* and *j>0*, eq. (52) becomes polynomial (51). For *λ(a)=1*, *λ(b)=-1* and *n=0*, the generalized eq. (52) becomes a sigmoidal one (5)-(7). For *λ(a)=0*, *λ(b)=λ(c)* and *j>0*, eq. (52) becomes eq. (8)-(9).

### 5.2. Generalized Gaussian and Bayesian fit

Returning to the regression analysis techniques it is possible to observe that the main assumption of the Gaussian method (eq. (10)) is that all points are treated as correct ones. This methodology causes potential misfits when the outliers exists. On the other hand the Bayesian fit assumes that all points can be potentially treated as outliers. However, generally one can propose the posterior probability function, analogically to eq. (10) and (13), which can connect both methods into a single one (Sivia & Skilling, 2003; Box & Tiao, 1968)

$$P_i = \varphi \int_{\sigma_{0i}}^{\infty} \mathcal{N}(E_i, \sigma_i^2) \frac{\sigma_{0i}}{\sigma_i^2} d\sigma_i + (1-\varphi) \mathcal{N}(E_i, \sigma_{0i}^2) \qquad (53)$$

where *N* is a normal (Gaussian) likelihood distribution and *φ* is the probability that data *E$_i$* is an outlier. It is the reason why the left-hand side of eq. (53) is a Bayesian distribution (same as eq. (13)) and the right-hand the Gaussian one (eq. (10)). This approach is called *Mixture of distributions* (Box & Tiao, 1968) or *The good-and-bad data model* (Sivia & Skilling, 2003). Thus the total posterior distribution for all *N* points can be found analogically as for eq. (14) and (23) as

$$S = \sum_{i=1}^{N} \ln \left\{ \frac{1}{\sigma_{0i}\sqrt{2\pi}} \left[ \varphi \frac{\sigma_{0i}^2}{R_i^2} \left( 1 - \exp\left(\frac{-R_i^2}{2\sigma_{0i}^2}\right) \right) + (1-\varphi) \exp\left(\frac{-R_i^2}{2\sigma_{0i}^2}\right) \right] \right\} \qquad (54)$$

where $R_i = Y_i - E_i$. Using the same reasoning of maximum likelihood method one can find the same solution as eq. (15), but with new weights, instead of $g_i$ (16), equal (Fornalski & Dobrzyński, 2010a)

$$g_i^* = \frac{1}{\sigma_{0i}^2} \cdot \frac{\exp\left(\frac{-R_i^2}{2\sigma_{0i}^2}\right) \cdot \left[\varphi \frac{\sigma_{0i}^2}{R_i^2}\left(2\frac{\sigma_{0i}^2}{R_i^2}+1\right)-1+\varphi\right]-2\varphi\frac{\sigma_{0i}^4}{R_i^4}}{\exp\left(\frac{-R_i^2}{2\sigma_{0i}^2}\right) \cdot \left(\varphi \frac{\sigma_{0i}^2}{R_i^2}-1+\varphi\right)-\varphi\frac{\sigma_{0i}^2}{R_i^2}} \tag{55}$$

The generalized method presented above is a good alternative for using Bayesian and Gaussian (least squares) fitting in the same time. One can see that for $\varphi=1$ the model became a Bayesian regression, while for $\varphi=0$ the model became a classical Gaussian (least squares) one. However, the mixed model works well just for $\varphi=0.05$ (Ekiz, 2002), because usually outlier points are a minority among all experimental data. Moreover, results obtained by weights $g_i^*$ (55) are very similar to ones obtained by $g_i$ (16), but with uncertainties often reduced by about 30% (Fornalski & Dobrzyński, 2010a).

### 5.3. Generalized Bayesian dose estimation method

For a generalization of presented Bayesian methods for dose estimation it is necessary to assume many parameters $\theta_i$ for each *i* type of radiation:

$$\theta_i = \frac{D_i}{\sum_{k=1}^{R} D_k} = \frac{D_i}{D_{total}} \tag{56}$$

and proper prior functions, $p(\theta_i)$. Priors should be assumed or established experimentally, as in eq. (31)-(34). When choosing, for the benefit of simplicity, the polynomial eq. (51), one can present $y_f$ dedicated for exact *i*-th dose as:

$$y_f(D_i) \equiv Y(D_i) = Y_0 + \sum_{i=1}^{R}\sum_{j=1}^{n} \lambda_{i,j} \left[\frac{\theta_i}{1-\theta_i}\left(\left(\sum_{k=1}^{R} D_k\right)-D_i\right)\right]^j \tag{57}$$

Each dose $D_i$ was written as a proper part of the total dose using eq. (56) and the reasoning used in eq. (45). Next, the likelihood function based on Poisson distribution (44) and $y_f(D_i)$ from eq. (57) can be written as:

$$L(D_i|\lambda_0, \lambda_1, \ldots, \lambda_{nR}, \theta_1, \theta_2, \ldots, \theta_R) \propto \left[y_f(D_i)\right]^u \times e^{-w\, y_f(D_i)} \tag{58}$$

Assuming, that all fitting parameters, $\lambda$, are given by their priors (including $\lambda_0=Y_0$), the posterior probability distribution for each dose equals

$$P(D_i) \propto \int \ldots \int L(D_i|\lambda_{0,\ldots,nR}, \theta_{1,\ldots,R}) \cdot \left(\prod_{j=0}^{nR} p(\lambda_j) d\lambda_j\right) \cdot \left(\prod_{k=1}^{R} p(\theta_k) d\theta_k\right) \tag{59}$$

which is the most general form of posterior probability for *R* types of radiation and *(n+1)R* number of fitting parameters.

## 6. DISCUSSION AND CONCLUSIONS

The presented paper is composed of four major parts: the first presents several forms of calibration curves, the second the robust Bayesian algorithm of fitting the proper curve to the experimental data, the third discusses five potential methods of dose estimation after mixed n+γ irradiation and the fourth presents certain generalized forms of methods presented before.

The robust Bayesian regression analysis method (Bayesian fit) is useful whenever outlier or scattered data points exists. The proposed method is useful to obtain fitting parameters of all potential curves, from simple linear (1) and quadratic (2) relationships to saturated functions (5)-(7) widely found in real conditions (Pacyniak et al., 2014; Dabrowski & Thompson, 1998). The algorithm allows for the calculation of the proper uncertainties of all parameters found while also providing relative plausibility of alternative models (fitted curves). The proposed algorithm is generally useful to find the best calibration curve for biological dosimetry. It was successfully used in practice on many occasions (Fornalski & Dobrzyński, 2009, 2010a, 2010b, 2011; Fornalski et al., 2010).

The five alternative statistical methods of the estimation of n+γ doses as well as proper chromosomal aberration frequencies were presented next. One can choose between the classical method, the classical method with probability distributions (called also a quasi-Bayesian method), two Bayesian methods and the Monte Carlo one. All methods have their own advantages as well as many inconveniences. It is not generally possible to indicate one method as the best, since this problem depends on the exact situation and expectations. Classical method is widely known and simple, but uses the exact value of the contribution of gamma dose in the total dose, $\theta$. When this parameter is known only as a proper probability distribution (prior function), $p(\theta)$, one can use enhanced classical method (called also quasi-Bayesian one) with probability distributions. Alternatively, it is possible to use the simplified Bayesian method, which allows for (as the only one) the use of the most inexact information about prior, given by eq. (34). The full Bayesian method proposed by Brame and Groer (2003) assumes prior function also for fitting parameters from the calibration curve. The last method, the Monte Carlo one, is an interesting computational alternative, most useful for statistical studies on irradiated groups of virtual cells. All presented methods were used in practice on real data (Pacyniak et al., 2014), see also Figs. 2-4.

Other, more general forms of methods presented before were also introduced. They can be particularly useful e.g. when the potential victim is irradiated by many types of radiation at the same time, not only gammas and neutrons. However, for practical cytogenetic dose estimation, certainly in an emergency scenario, the generalized methods could also be useful for separating gradient type inhomogeneous exposures, e.g. irradiation of different fractions of the body.

The presented statistical methods can be applied into the computational algorithm and be used in cytogenetic analysis. However, biologists and cytogeneticists less experienced in the use of the mathematical methods may need some help of programmers, mathematicians or physicists. Nevertheless, the methods are usually presented in easy-to-use forms, which can be applied even as Excel's formulas.

The presented paper should be treated as a methodology and statistical guide. The description of detailed application of presented methods to dedicated experimental data was intentionally omitted, except for some examples in Figs. 1-4. The application of all methods to real data is the subject of other studies (Pacyniak et al., 2014).

**ACKNOWLEDGEMENTS**

This work was supported by The National Centre for Research and Development (NCBiR), Poland, under the strategic research project: *"Technologies supporting development of safe nuclear power engineering"*, research task no. 6: *"Development of nuclear safety and radiological protection methods for the nuclear power engineering's current and future needs"*, subject no. 8: *"Implementation of the dicentric assay for biological dosimetry following accidental exposure to fission neutrons"*.

The author wishes to thank Dr Maria Kowalska from Central Laboratory for Radiological Protection (Poland) for the inspiration, careful review of the manuscript as well as many valuable remarks and discussions about the cytogenetics. Additional thanks for Dr Y. Socol, Prof. P.G. Groer, Dr R.S. Brame, Prof. L. Dobrzyński, Ms. I. Pacyniak and Ms. K. Buraczewska for additional comments.